# Dynamics of Self-injected Electron Bunches at their Acceleration by Laser Pulse in Plasma

*D.S. Bondar[2], I.P. Levchuk[1], V.I. Maslov[1], S. Nikonova[2], I.N. Onishchenko[1]*

[1] *NSC Kharkov Institute of Physics & Technology 61108 Kharkov, Ukraine*
[2] *Karazin Kharkiv National University 61022, Kharkov, Ukraine*
e-mail: vmaslov@kipt.kharkov.ua

Dynamics of self-injected electron bunches has been demonstrated by numerical simulation in blowout or bubble regime at self-consistent change of mechanism of electron bunch acceleration by plasma wakefield, excited by a laser pulse, to additional accelerating mechanism of electron bunch by plasma wakefield, excited by self-injected electron bunch. Two methods of acceleration: by one laser pulse and by short chain of two laser pulses have been numerically simulated. Possibility and advantages of injection of train of laser pulses have been considered. Ways of separation of laser pulse in shaped train of laser pulses have been presented. The causes of defocusing of electron bunches at wakefield excitation by laser pulse in plasma in blowout or bubble regime have been considered.

## 1. INTRODUCTION

Laser-plasma-based accelerators are of great interest now [1 – 16]. Successful experiments on laser wakefield acceleration of charged particles in the plasma have confirmed the relevance of this method of acceleration [2 – 7, 12]. The formation of electron bunches with small energy spread was demonstrated at intense laser–plasma interactions [17 - 19]. Processes of a self-injection and colliding injection of electrons and their acceleration have been experimentally studied in a laser-plasma accelerator [20].

The problem at laser wakefield acceleration is that laser pulse quickly destroyed because of its expansion. One way to solve this problem is the use of a capillary as a waveguide for laser pulse. The second way to solve this problem is to transfer its energy to the electron bunches which as drivers accelerate witness. A transition from a laser wakefield accelerator to plasma wakefield accelerator can occur in some cases at laser-plasma interaction [21]. Self-injected electron bunches play an important role in the interaction of intense laser pulse with the plasma.

The main aim of this paper is the kinetic numerical simulation of the dynamics of self-injected and accelerated electron bunches by wakefield in blowout or bubble regime, excited by the laser pulse in the plasma.

## 2. DYNAMICS OF SELF-INJECTED ELECTRON BUNCHES

Fully relativistic electromagnetic 2.5D particle–in–cell simulation was carried out by the UMKA2D3V code (Institute of Computational Technologies) [22]. A computational domain (x, y) has a rectangular shape with the dimensions: $0 < x < 800\lambda$ and $0 < y < 50\lambda$, where $\lambda$ is the laser pulse wavelength, $\lambda = 80$ μm. The computational time interval is $\tau = 0.05$, the number of particles per cell is 8 and the total number of particles is $15.96 \times 10^6$. The period of the laser pulse is $t_0 = 2\pi/\omega_0$, where $\omega_0$ is the laser frequency. The laser pulse is injected normally into uniform plasma from the left boundary. In transversal $y$ direction, the boundary conditions for particles, electric and magnetic fields are periodic. The plasma density is $n_0 = 0.01016 n_c = 1.8 \times 10^{19}$ cm$^{-3}$, where the critical plasma density $n_c = m_e\omega_0^2/(4\pi e^2)$, $m_e$ is the electron mass, $e$ is the electron charge. The laser pulse is Gauss-type in longitudinal direction. The pulse has a Gaussian profile in the transverse direction. The longitudinal and transverse dimensions of the laser pulse are selected to be shorter than the plasma wavelength. The length at half maximum equals $2\lambda$ and width at half maximum equals $8\lambda$. The simulations were performed for the peak normalized laser field strength, $a_0 = eE_{x0}/(m_e c\omega_0) = 5$, where $E_{x0}$ is the electric field amplitude, $c$ is the speed of light. The latter corresponds to the peak laser intensity $I = 5.3 \times 10^{19}$ W/cm$^2$. Below coordinates $x$ and $y$, time $t$, electric field amplitude $E_x$ and electron plasma density $n_0$ are given in dimensionless form in units of $\lambda$, $2\pi/\omega_0$, $m_e c\omega_0/(2\pi e)$, $m_e\omega_0^2/(16\pi^3 e^2)$, correspondingly.

We consider both cases: injection of single laser pulse and injection of a short train of two laser pulses. 1st reason for using a short train of laser pulses, that after the 1st bubble there is wake it is useful for increase efficiency and charge (current) of accelerated electrons to enhance its by next laser pulse and to use for acceleration of additional electron bunches.

2-nd reason for using a short train of laser pulses, that since the best results on electron bunch acceleration by laser pulse have been achieved not at the maximum parameters of the laser pulse, it is advantageous to convert the laser pulse with the maximum parameters in a train of several pulses and to receive increased current of accelerated electrons.

It is advantageous to use a train of laser pulses because in this case it is easier to get a large transformation ratio TR>>1. TR can be defined as the ratio of the maximum wake perturbation after the driver to the maximum wake perturbation within the driver. It characterizes the maximum achievable energy of accelerated electrons at a given energy of driver. I.e. at larger TR, the more energy can be obtained by the accelerated electrons. It is known that for a typical driver TR≤2.

To obtain a large TR>>1, i.e. to increase transformation ratio of laser pulse energy into energy of accelerated electron bunches it is necessary to create a

shaped on intensity $I_i$ train of laser pulses in accordance with:

$I_1:I_2:I_3: \ldots = 1:3:5: \ldots$ (drivers are located approximately in $1.5\lambda$)

or

$I_1:I_2:I_3: \ldots = 1:2:3: \ldots$ (drivers are located approximately in $\lambda$)

or

$I_1:I_2:I_3: \ldots = 1:5:9: \ldots$ (drivers are located approximately in $1.5 \lambda$)

With modern technology, this is easily solved in experiments. For example, from one powerful laser pulse in experiment, it is possible to obtain a profiled short train of laser pulses by splitting intense laser pulse (see Fig. 1).

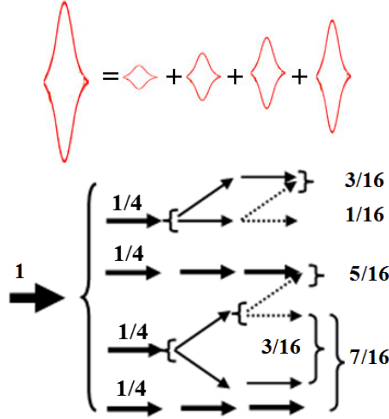

Fig. 1. Intense laser pulse splitting into train of micropulses, spaced by $1.5\lambda$, $\lambda$ is fundamental plasma wavelength

In this case, the ratio of laser pulse intensities equals
$I_1:I_2:I_3:I_4=1/16:3/16:5/16:7/16.$
The latter is similar to the following
$I_1:I_2:I_3:I_4=1:3:5:7.$
In this case, the transformation ratio is approximately equal to TR≈8.

The second method of producing a profiled short train of laser pulses by splitting intense laser pulse is following (see Fig. 2). In this case, the ratio of laser pulse intensities equals
$I_1:I_2:I_3:I_4:I_5 =1/16:1/8:3/16:1/4:5/16.$
The latter is similar to the following
$I_1:I_2:I_3:I_4:I_5 =1:2:3:4:5.$
In this case, the transformation ratio is approximately equal to TR≈5π.

Large TR can be obtained by shaping one powerful laser pulse.

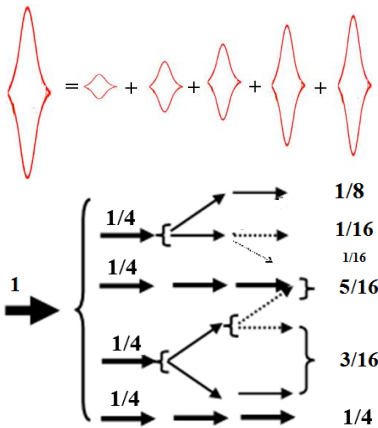

Fig. 2. Intense laser pulse splitting into train of micropulses, spaced by $\lambda$, $\lambda$ is fundamental plasma wavelength

We consider the following scenario for the acceleration of electrons by a laser pulse in a plasma. First, the laser pulse (or laser pulses) excites the wakefield. Then the 1st self-injected and accelerated bunches become drivers and together with the laser pulse (or laser pulses) accelerate the last self-injected bunches. Over time, the laser pulse (or pulses) is destroyed. Later, the driver-bunches are also destroyed and bubbles disappear. Let's consider some features of dynamics of bunches at this chain of change of drivers.

In 1st front bubble remains partially filled with plasma electrons. Therefore, the radial focusing force of the bubble is highly inhomogeneous (see Fig. 3).

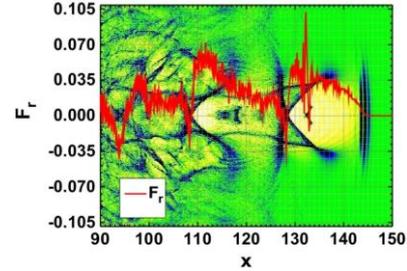

Fig. 3. Wake perturbation of plasma electron density $n_e$ and off-axis transverse wake force $F_\perp \triangleright E_y - B_z$ (red line) excited by one laser pulse

This leads to an increase of amplitude of the betatron oscillations of the electron bunches.

The increase of the amplitude of the betatron oscillations of the electron bunches in the 1st half of bubble is also determined by the fact that bunches are decelerated there. Hence, the defocusing force of their space charge increases.

Self-cleaning (similar to [23, 24]) of the decelerated 1st bunch in the 2nd bubble in the case of injection of one laser pulse or 1st bunch in the 3rd bubble in the case of injection of two laser pulses occurs due to growth of amplitude of betatron oscillations with decreasing focusing radial force $F_r(z)$ of bubble to its 1st front and at increase of defocusing force of the volume charge of the bunch at its deceleration in front of the bubble.

If the focusing radial force of the 1st bubble decreases slowly and it breaks sharper just near the leading edge of the bubble, where 1st bunch reaches the laser pulse, 1st bunch strongly defocuses only near the leading edge of the bubble (see Figs. 4, 5).

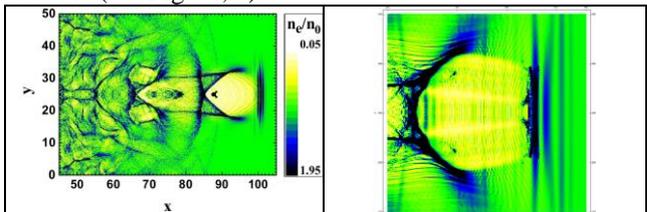

| Fig. 4. Formation of 1st self-injected electron bunch in 1st bubble, excited by one laser pulse | Fig. 5. Expansion of 1st self-injected electron bunch in 1st bubble at its reaching laser pulse and expansion of 2nd self-injected electron bunch in |

| | 1st bubble at t = 490$t_0$ |

Witness in 1st bubble, until it reaches a high energy due to acceleration and if it is self-injected too far from the axis from the walls of the bubble, it greatly expands due to the expansion of betatron oscillations in the decreased in longitudinal direction focusing radial force of bubble (see Figs. 5, 6).

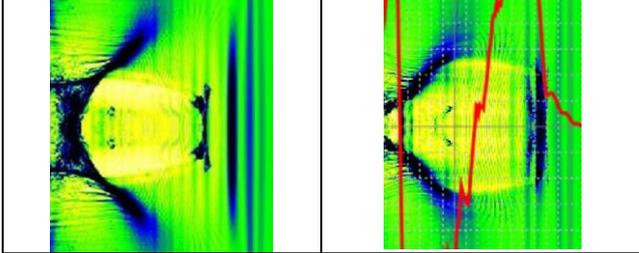

| Fig. 6. 2nd self-injected electron bunch in 1st bubble at t = 435$t_0$ | Fig. 7. Expansion of 1st self-injected electron bunch in 1st bubble at its reaching laser pulse and expansion of 2nd self-injected electron bunch in 1st bubble at t = 490$t_0$ |

At further acceleration of the 2nd bunch in the 1st bubble, the bunch is increasingly compressed to the axis due to a decrease of amplitude of the betatron oscillations at increasing $F_r(z)$ (see Figs. 5, 7).

Thus, the formation of four electron bunches is observed (two in each bubble) (1st is the driver, 2nd is the witness) in the 1st and 2nd bubbles (see Figs. 4, 5) in the case of a single pulse injection or in 1st and 3rd bubbles in the case of two pulses injection (see Figs. 5, 8).

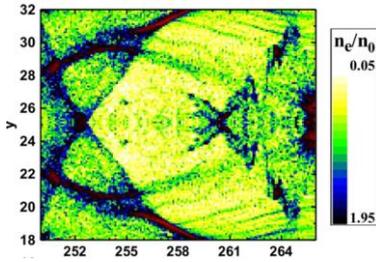

Fig. 8. Formation of 1st and 2nd self-injected electron bunches in 3rd bubble, excited by two laser pulses

Then when the 2nd pulse is destroyed, a bunch can be formed in the 2nd bubble in the case of two pulses injection (see Fig. 9).

Witness, which has become a driver, continues to form the bubble, since its charge density is larger than the plasma density $n_b > n_{0e}$, although the laser has been destroyed (see Fig. 7).

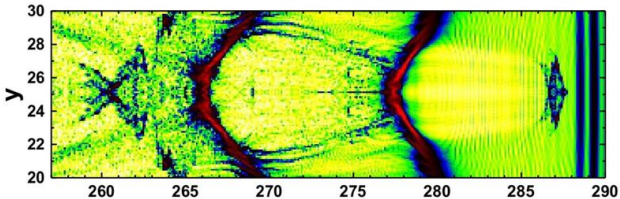

Fig. 9. Formation of self-injected electron bunch in 2nd bubble, excited by two laser pulses

Then after some time the bunch is destroyed and, of course, bubble disappears too.

At certain parameters, the ratio of the bunch density to the plasma density is equal to $n_b \approx 6 n_{0e}$. It roughly corresponds to the ratio of the bubble length to the length of steeping.

If we assume that with a change of the plasma density $n_{0e}$, the dimensions of the self-injected bunch are the same fraction of the bubble length, then the bunch charge is proportional to

$$n_{0e}\lambda^3 = n_{0e}(\pi c/\omega_{pe})^3 \sim 1/\sqrt{n_{0e}}.$$

Therefore, to obtain a large charge of a bunch, it is impossible to increase the plasma density. It is necessary to increase the plasma density $n_{0e}$ if we want to obtain bunches with small dimensions. And if we want to obtain bunch with a large charge $q_b$, we must decrease the plasma density $n_{0e}$.

Such a large density of the bunch $n_b \approx 6 n_{0e}$ can be maintained for a long time.

## 3. CONCLUSION

Dynamics of self-injected electron bunches has been demonstrated by numerical simulation in blowout or bubble regime at self-consistent change of mechanism of electron bunch acceleration by plasma wakefield, excited by a laser pulse, to additional accelerating mechanism of electron bunch acceleration by plasma wakefield, excited by self-injected electron bunch.

Two methods of acceleration: by one laser pulse and by short chain of two laser pulses have been numerically simulated.

Possibility and advantages of injection of train of laser pulses have been considered.

Ways of separation of laser pulse in shaped train of laser pulses have been presented.

The causes of defocusing of electron bunches at wakefield excitation by laser pulse in plasma in blowout or bubble regime have been considered.

## ACKNOWLEDGEMENTS

We acknowledge financial support from the National Research Fund of Ukraine "Support for research of leading and young scientists" grant "Transport of electron/positron bunches at high-gradient acceleration by electromagnetic fields excited in dielectric structures or plasma by a high power electron bunches and an intense laser pulse" 193/02.2020.

## ДИНАМИКА САМОИНЖЕКТИРУЕМЫХ ЭЛЕКТРОННЫХ СГУСТКОВ ПРИ ИХ УСКОРЕНИИ ЛАЗЕРНЫМ ИМПУЛЬСОМ В ПЛАЗМЕ

*Д.С.Бондарь, И.П.Левчук, В.И.Маслов, С.Никонова, И.Н.Онищенко*

Численным моделированием продемонстрирована динамика самоинжектированных электронных сгустков в нелинейном режиме опрокидывания при самосогласованном изменении механизма ускорения электронов полем кильватерной волны, возбуждаемой лазерным импульсом в плазме, на дополнительный механизм ускорения электронов полем кильватерной волны, возбуждаемой самоинжектированным электронным сгустком в плазме. Численным моделированием исследованы два способа ускорения: одним лазерным импульсом и короткой цепочкой из двух лазерных импульсов. Рассмотрены возможность и преимущества инжекции последовательности лазерных импульсов. Представлены способы трансформации лазерного импульса в профилированную последовательность лазерных импульсов. Рассмотрены причины дефокусировки электронных сгустков при возбуждении кильватерного поля лазерным импульсом в плазме в нелинейном режиме опрокидывания.

## ДИНАМІКА САМОІНЖЕКТОВАНИХ ЕЛЕКТРОННИХ ЗГУСТКІВ ПРИ ЇХ ПРИСКОРЕННІ ЛАЗЕРНИМ ІМПУЛЬСОМ В ПЛАЗМІ

*Д.С.Бондарь, І.П.Левчук, В.І.Маслов, С.Ніконова, І.М.Онищенко*

Чисельним моделюванням продемонстрована динаміка самоінжектованих електронних згустків в нелінійному режимі перекидання при самоузгодженій зміні механізму прискорення електронів полем кільватерної хвилі, яка збуджується лазерним імпульсом в плазмі, на додатковий механізм прискорення електронів полем кільватерної хвилі, яка збуджується самоінжектованим електронним згустком в плазмі. Чисельним моделюванням досліджені два способи прискорення: одним лазерним імпульсом і коротким ланцюжком з двох лазерних імпульсів. Розглянуто можливість і переваги інжекції послідовності лазерних імпульсів. Представлені способи трансформації лазерного імпульсу в профільовану послідовність лазерних імпульсів. Розглянуто причини дефокусування електронних згустків при збудженні кільватерного поля лазерним імпульсом в плазмі в нелінійному режимі перекидання.